\documentclass{WileyMSP-template}
\usepackage[utf8]{inputenc}

\usepackage{gensymb}
\usepackage{amsmath}

\usepackage{graphicx}
\usepackage{subcaption}
\usepackage[dvipsnames]{xcolor}
\usepackage{textcomp}
\usepackage{booktabs}
\begin{document}

\pagestyle{fancy}
\rhead{\includegraphics[width=3cm]{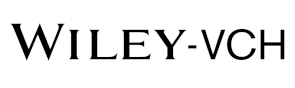}}

\title{Deterministic nanofabrication of quantum dot-circular Bragg grating resonators with high process yield using in-situ electron beam lithography} {Deterministic nanofabrication of quantum dot-circular Bragg grating resonators with high process yield using in-situ electron beam lithography}

\maketitle


\author{Avijit Barua\textsuperscript{*}},
\author{Kartik Gaur},
\author{Léo J. Roche},
\author{Suk In Park},
\author{Priyabrata Mudi\textsuperscript{*}},
\author{Sven Rodt},
\author{Jin-Dong Song} and
\author{Stephan Reitzenstein\textsuperscript{*}}

\begin{affiliations}
A. Barua, K. Gaur, L. J. Roche, Dr. P. Mudi, Dr. S. Rodt, Prof. Dr. S. Reitzenstein\\
Institut für Physik und Astronomie, Technische Universität Berlin, Hardenbergstraße 36, D-10623 Berlin, Germany\\
Email Address: avijit.barua@tu-berlin.de, mudi@tu-berlin.de, stephan.reitzenstein@physik.tu-berlin.de\\
Dr. S. I. Park, Prof. Dr. J-D. Song\\
Korea Institute of Science and Technology (KIST), Seoul, Republic of Korea

\end{affiliations}



\begin{abstract}

The controlled integration of quantum dots (QDs) as single-photon emitters into quantum light sources is essential for the implementation of large-scale quantum networks. In this study, we employ the deterministic in-situ electron-beam lithography (iEBL) nanotechnology platform to integrate individual QDs with high accuracy and process yield into circular Bragg grating (CBG) resonators. Notably, CBG devices comprising just 3 to 4 rings exhibit photon extraction efficiencies comparable to those of structures with more rings. This facilitates faster fabrication, reduces the device footprint, and enables compatibility with electrical contacting. To demonstrate the scalability of this process, we present results of 95 optically active QD-CBG devices fabricated across two lithography sessions. These devices exhibit bright, narrow-linewidth single-photon emission with excellent optical quality. To evaluate QD placement accuracy, we apply a powerful characterization technique that combines cathodoluminescence (CL) mapping and scanning electron microscopy. Statistical analysis of these devices reveals that our iEBL approach enables high alignment accuracy and a process yield of over $>90\%$ across various CBG geometries. Our findings highlight a reliable route toward the scalable fabrication of high-performance QD-based single-photon sources for use in photonic quantum technology applications.
\end{abstract}


\section{Introduction}

Bright single-photon sources are a fundamental building block for quantum technologies, particularly in the fields of quantum communication~\cite{arakawa2020, Maring2024} and photonic quantum computing \cite{wang2020integrated}. They are essential for enabling key functionalities, such as secure information transfer in photonic quantum networks \cite{Lodahl_2018}, as well as the implementation of scalable quantum computing architectures. Key to the development of such systems are single-photon sources that exhibit high brightness and spectral purity and can be fabricated in a scalable manner with a high yield \cite{NemotoKae2014, Heindel2023, aghaeerad2025}. Among the available solid-state quantum emitters, self-assembled InGaAs quantum dots (QDs) stand out as highly promising candidates for on-demand single-photon emission \cite{Michler2000, Senellart2017}. Their discrete atom-like energy levels enable deterministic single-photon emission, offering key advantages over probabilistic sources such as those based on spontaneous parametric down-conversion \cite{baghdasaryan2023enhancing}. These properties provide a robust foundation for on-chip quantum photonic integration, making QDs an attractive platform for developing compact, scalable quantum devices.

Despite major advantages in QD research and device engineering, the stochastic nature of self-assembled QD nucleation remains a major obstacle to scalable quantum photonic integration. In addition, uncertainty in the emission energy, driven by the fluctuation in QD size, composition, and local strain, further constrains their integration with a suitable photonic cavity, which requires a precise spatial and spectral match with their photonic modes. These mismatches undermine the deterministic coupling to photonic cavities, thereby limiting device yield, reproducibility, and scalability. Although site-controlled growth strategies show promise \cite{Gaur2025MQT}, overcoming the limitations of self-assembled QDs remains crucial for enabling the scalable fabrication. In this context, high-performance devices are primarily produced using deterministic integration techniques that combine low-temperature photoluminescence (PL) mapping with marker-based electron beam lithography (EBL) at room temperature \cite{Sapienza2015, Liu2017, Rickert2025}. However, these workflows require repetitive cooldown-warmup cycles and alignment procedures and are prone to cumulative drift, spatial misalignment, and fabrication imperfections. Notably, spatial misalignment that places the QD in close proximity to etched surface or cavity sidewalls can critically impair the coupling efficiency and significantly degrade the emitter's optical properties, including brightness, coherence, and single-photon purity, due to increased surface-related non-radiative recombination and environmental decoherence \cite{Zhao2024}. These limitations pose a fundamental challenge to wafer-scale or industrial deployment. To overcome these restraints, marker-free, deterministic integration schemes that rely on emitter selection and device fabrication in a single cryogenic session have emerged as critical enablers of scalable quantum technology. In-situ approaches, including cryogenic optical lithography \cite{dousse2008controlled} and in-situ EBL (iEBL) \cite{gschrey2013situ}, allow nano-photonic structures to be directly written around pre-selected emitters with spatial precision as good as sub-50 nm \cite{gschrey2013situ}. Among these, iEBL uniquely combines cathodoluminescence (CL)-based QD identification with flexible, high-resolution lithography \cite{madigawaDonges2024, Gshrey2015, Schnauber2018, Rodt2021}, enabling streamlined, high-throughput fabrication while eliminating thermal drift and maintaining high optical quality. 

Building upon a systematic study of the deterministic integration of QDs into simple mesa structures \cite{madigawaDonges2024}, in which the alignment accuracy of marker-based EBL and iEBL was compared, we focus here on a comprehensive investigation of the integration accuracy and process yield of iEBL for embedding QDs into circular Bragg grating (CBG) resonators \cite{nawrath2023bright, Setthanat2024}, which have high application potential. CBGs offer highly directional vertical light extraction, which is a key requirement for coupling single photons in free space or fiber-based quantum networks \cite{Jeon2022fibercoupling, bremer2022fiber, schwab2022coupling}. If optimized for pronounced light-matter interaction in the cavity quantum electrodynamics regime, they also feature strong Purcell enhancement of spontaneous emission by more than a factor of over twenty \cite{Rickert2025}, which can help to improve the photon indistinguishability \cite{liu2018high}. Importantly, the broad spectral tolerance of CBGs allows for more relaxed constraints on the spectral alignment between the QD and the cavity mode if compared, for instance, to micropillar cavities \cite{reitzenstein2010quantum}. 

We fabricated a total of 103 QD-CBG devices across two fabrication sessions, achieving 95 optically active devices with an over 90\% processing yield.  Specifically, we have systematically investigated the impact of CBG geometry by fabricating and characterizing structures ranging from 1-ring to 5-ring CBGs. This comprehensive ring-number-dependent study evaluates the optical performance and emission quality of various designs, providing experimental insights into the trade-offs between fabrication complexity, brightness, and spectral characteristics. Notably, CBG devices with only 3-4 rings achieve photon extraction efficiencies comparable to those of devices with more rings, while offering advantages such as reduced fabrication time, a smaller device footprint, and improved compatibility with electrically contacted architectures.

Within a statistical analysis of the alignment accuracy, we combined two-dimensional (2D) Gaussian fitting of CL emission maps with scanning electron microscopy (SEM) of a total of 60 deterministically fabricated devices that met the emission spots described by the 2D Gaussian. 35 devices with asymmetric CL profiles were excluded to avoid bias from non-Gaussian fits. This analysis reveals high alignment accuracy across varying geometries and fabrication runs. The best-performing devices achieved a maximum PEE of 68\%, with emission linewidths limited by spectrometer resolution (30 µeV). Second-order correlation measurement confirms highly pure single-photon emission with $g^{(2)}(0) = 0.011 \pm 0.002$, with consistent structural quality. Overall, our findings demonstrate that cryogenic iEBL is a high-yield, scalable approach for the deterministic integration of QDs into photonic architectures, enabling a practical pathway to wafer-scale fabrication with high precision in a single cryogenic step.

\section{Design and fabrication of photonic structures}

This work employs a photonic architecture based on CBG resonators. These resonators are designed and numerically optimized to maximize the photon extraction efficiency (PEE), matching the numerical aperture (0.8) of the optical setup. Each CBG consists of a central GaAs mesa that embeds a single InGaAs QD, which is surrounded by concentric etched GaAs rings. The device design also includes a low-index SiO\textsubscript{2} spacer and an Au back reflector. While the current design is compatible with Purcell enhancement of QD emission, optimization and detailed investigation of this aspect were not the focus of this study. The CBG designs were systematically varied throughout the fabrication process, ranging from 1-ring to 5-ring gratings.

We carried out numerical optimizations to determine the optimal device geometry for maximum PEE by employing a Bayesian optimization algorithm. The PEE values were calculated using FEM simulations in JCMsuite solver \cite{Burger_JCM, JCMsuite}, based on a 2D model with cylindrical symmetry that incorporates the flip-chip device configuration with a backside gold mirror. The QD emitter was modeled as a point-like classical dipole radiating at a wavelength of 895 nm. The optimization explored multiple geometric parameters, including the mesa diameter, ring width, ring gap, and etching depth.  Each parameter was treated as free for optimization within physically relevant ranges. Simulations were conducted for a total of 5 device variants, starting with a 1-ring and adding rings incrementally up to a 5-ring configuration. To maintain comparability across all configurations, the ring width and ring gap were kept constant within each design to ensure periodicity in the grating design. For each geometry, the PEE was first maximized with the emitter placed in the center of the mesa to ensure a near-Gaussian far-field emission profile. Subsequently, the impact of a lateral QD displacement relative to the mesa center was systematically analyzed to assess the alignment tolerance and the device performance robustness.
These devices are spectrally targeted at 895 nm to ensure compatibility with the atomic vapor quantum memories, aligning with the Cs-D1 transition \cite{MaassBarua2025}. This makes the scalable approach directly relevant for quantum network applications. The optimized CBG geometries used for deterministic device fabrication are summarized in Table 1.

\begin{table}[h!]
\centering
\caption{Geometrical parameters and simulated key parameters of CBG devices with varying ring numbers.}
\begin{tabular}{|c|c|c|c|c|}
\hline
\textbf{Design} & \textbf{Mesa Diameter (nm)} & \textbf{Ring Width (nm)} & \textbf{Gap Width (nm)} & \textbf{PEE$_{max}$ (\%)}  \\
\hline

CBG with 1 ring   & 1242    & 215   & 450  & $85 \pm 1$  \\
CBG with 2 rings  & 1238    & 210   & 450  & $87 \pm 1$  \\
CBG with 3 rings  & 1238    & 205   & 443  & $84 \pm 1$  \\
CBG with 4 rings  & 1239    & 215   & 450  & $87 \pm 1$  \\
CBG with 5 rings  & 1238    & 210   & 450  & $87 \pm 1$  \\
\hline
\end{tabular}
\end{table}
Optimized geometrical parameters to maximize PEE and to ensure a near-Gaussian far-field emission profile suitable for efficient fiber coupling.
\vspace{0.5 cm}

\begin{figure}
 \centering
  \includegraphics[width=1\linewidth]{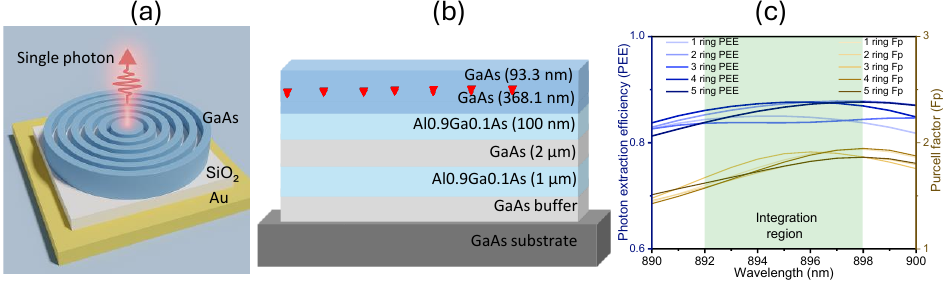}
  \caption{
  (a) Final hybrid flip-chip bonded device architecture with SiO\textsubscript{2}/Au mirror stack. Conceptual schematic diagram of the single-photon emission process from a QD embedded in a CBG resonator.
  (b) Layer structure of the epitaxially grown heterostructure comprising a GaAs membrane and sacrificial Al\textsubscript{0.9}Ga\textsubscript{0.1}As layers. QDs are indicated by red triangles.
  (c) FEM simulation results showing photon extraction efficiency and Purcell factor for CBGs with 1 to 5 rings, highlighting the $(895 \pm 3)$ nm integration window.  
  }
  \label{fig:1}
\end{figure}

The applied photonic design and the corresponding simulation results are presented in Fig.~\ref{fig:1}. Panel (a) illustrates the hybrid CBG device fabricated using flip-chip bonding and CBG patterning, which indicates single-photon emission from a QD embedded in the CBG resonator on top of a reflective substrate.
Fig.~\ref{fig:1}(b) depicts the epitaxially grown heterostructure, consisting of a GaAs membrane and sacrificial layers consisting of Al$_{0.9}$Ga$_{0.1}$As. 
Fig.~\ref{fig:1}(c) shows the wavelength-dependent PEE and Purcell factor ($F_p$) obtained via FEM-modelling for CBGs with 1 to 5 concentric rings. A high PEE value greater than 80\% is maintained within the $(895 \pm 5)$ nm spectral range, which defines the target window for QD device integration. Having the atomic quantum memory application in mind, the CBG parameters (Table 1) are also optimized to maximize PEE without significant changes $F_p$ in the range of $\mu = (1.81 \pm 0.13)$, ensuring that the lifetime-limited homogeneous linewidth of the emission remains as narrow as possible for efficient coupling to the Cs-D1 transition while maintaining high photon extraction efficiency \cite{MaassBarua2025}.

The device fabrication begins with the epitaxial growth of InGaAs QDs on a GaAs wafer via molecular beam epitaxy (MBE) \cite{UdoPohlMBE}. The QD layer is positioned 93.3 nm beneath the surface of a 461.4 nm thick GaAs layer. Two sacrificial AlGaAs layers are incorporated underneath the membrane to facilitate the subsequent release and processing steps. After growth, the sample is bonded via Au-Au thermo-compression bonding onto a carrier chip, resulting in a gold (Au) back reflector. A thin SiO\textsubscript{2} dielectric spacer layer is introduced between the GaAs membrane and the Au mirror to form a reflective and thermally stable hybrid backplane \cite{WangJWPan2019, Jinliu2019}. To execute deterministic patterning, the bonded sample is cooled to 20 K inside a state-of-the-art EBL system (Raith eLine Plus), which is equipped with a helium-flow cryostat and a CL unit for iEBL processing. Before mounting, a 300 nm-thick layer of CSAR-6200:13 resist is spin-coated onto the sample and baked at 150°C for one minute. Consequently, CL mapping is used to identify spectrally isolated QDs emitting in the spectral range of $(895 \pm 3)$ nm. Based on the recorded CL maps, proximity-corrected electron beam patterns for CBG structures are in-situ written into the resist \cite{Rodt2021}, precisely aligned to the emission center of each selected QD. Since all the steps are executed within the same cryogenic session, a simple and straightforward process flow is ensured, eliminating the need for any post-alignment procedures. After exposure, the sample is removed from the EBL system to develop the resist and perform pattern transfer into the GaAs membrane. This is done using inductively coupled plasma reactive ion etching (ICP-RIE) with Cl\textsubscript{2}/Ar chemistry in a clean room environment. The etching extends through the entire 461.4 nm membrane, producing well-defined features with high aspect ratios that preserve the intended CBG geometries. This single-session, marker-free iEBL process eliminates conventional sources of spatial misalignment and enables high-throughput batch fabrication of QD–CBG devices with excellent structural fidelity and uniformity across multiple samples and design variants.

\section{Scalability and process uniformity}

The primary objective of the fabrication strategy employed in this work is to develop a reproducible and scalable approach for integrating QDs into photonic nanostructures with high spatial and structural fidelity. To evaluate the alignment accuracy associated with this approach, extensive batch fabrication of QD-CBG devices was carried out across multiple cool-downs, substrate regions, and design variants. All devices were processed using the aforementioned marker-free iEBL workflow to ensure consistent alignment and pattern transfer throughout the fabrication pipeline.

The fabrication phase consisted of 103 QD devices across two sessions, incorporating CBG structures ranging from 1 to 5 concentric rings. These arrays were patterned over \(20\times20~\mu\mathrm{m}^2\) fields, demonstrating the compatibility of the process with dense layouts and complex geometries. 
The process consistently yielded fully formed structures with intact morphology, characterized by the preservation of structure height, ring width, and gap width without collapse, deformation, or etch-induced erosion and precise alignment. The initial µPL measurements showed that 95 devices were optically active, resulting in the process yield exceeding 90\%. Maintaining such morphological integrity is critical to ensure both the optical functionality and mechanical robustness of the resonator, especially when fabricated at high density over extended chip areas.

The post-processing evaluation started with high-resolution SEM imaging, which confirmed consistent pattern transfer and structural uniformity across all device designs. The mesa and ring dimensions were within 20-30 nanometers of their target values, and the etched sidewalls exhibited smooth vertical profiles and surface texture. These results highlight the effectiveness of the proximity-effect correction and the mechanical stability of the cryogenic iEBL setup. Using CL-based emitter selection has enabled the precise and deterministic targeting of spatially and spectrally isolated QDs throughout the entire field of view. The integration process of QD–CBG exhibited consistent performance across extended fabrication sessions.

\begin{figure}
 \centering
  \includegraphics[width=1\linewidth]{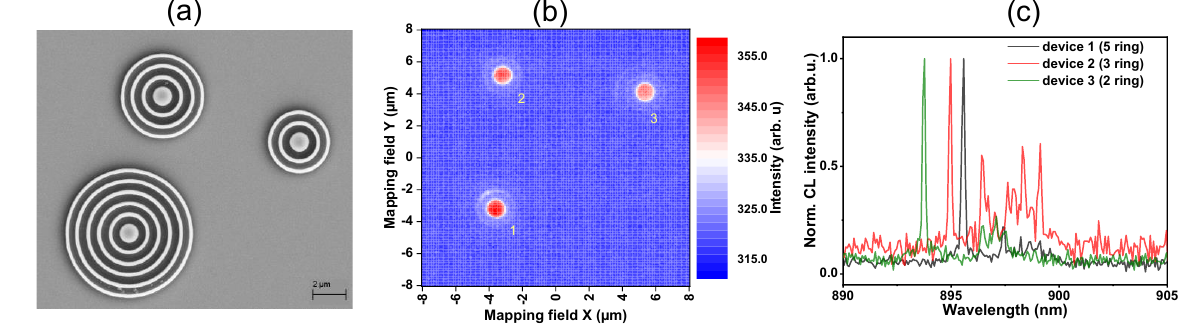}
  \caption{Structural and optical overview of a representative 20 × 20 µm² device field fabricated via the iEBL process. (a) High-resolution SEM image displays three exemplary QD-CBG structures with increasing ring numbers (2, 3, and 5 rings) in a single mapping field. (b) The CL intensity map, acquired from the corresponding field at 20 K, with consistent photon emission from the integrated QD. (c) Normalized CL spectra from three devices to confirm spectral reproducibility and successful spectral targeting during iEBL integration. 
}
  \label{fig:2}
\end{figure}

As illustrated in Fig.~\ref{fig:2}, the structural and optical characteristics of a representative \(20 \times 20~\mu\mathrm{m}^2\) field are presented. Panel~(a) shows an SEM image of QD-CBG devices with 2, 3, and 5 Bragg rings (more SEM images of different designs are provided in the supporting information (SI)). The high etch precision and dimensional consistency reflect the robustness of the fabrication process. Fig.~\ref{fig:2}(b) displays the corresponding CL intensity map, acquired at 20 K, revealing strong, localized emission at each cavity center. Fig.~\ref{fig:2}(c) presents a representative CL spectrum centered at the target wavelength of $(895 \pm 3)$ nm from all 3 devices. The sharp emission peak with minimal background indicates excellent optical quality. 

\section{Alignment accuracy characterization}

\begin{figure}
 \centering
  \includegraphics[width=1\linewidth]{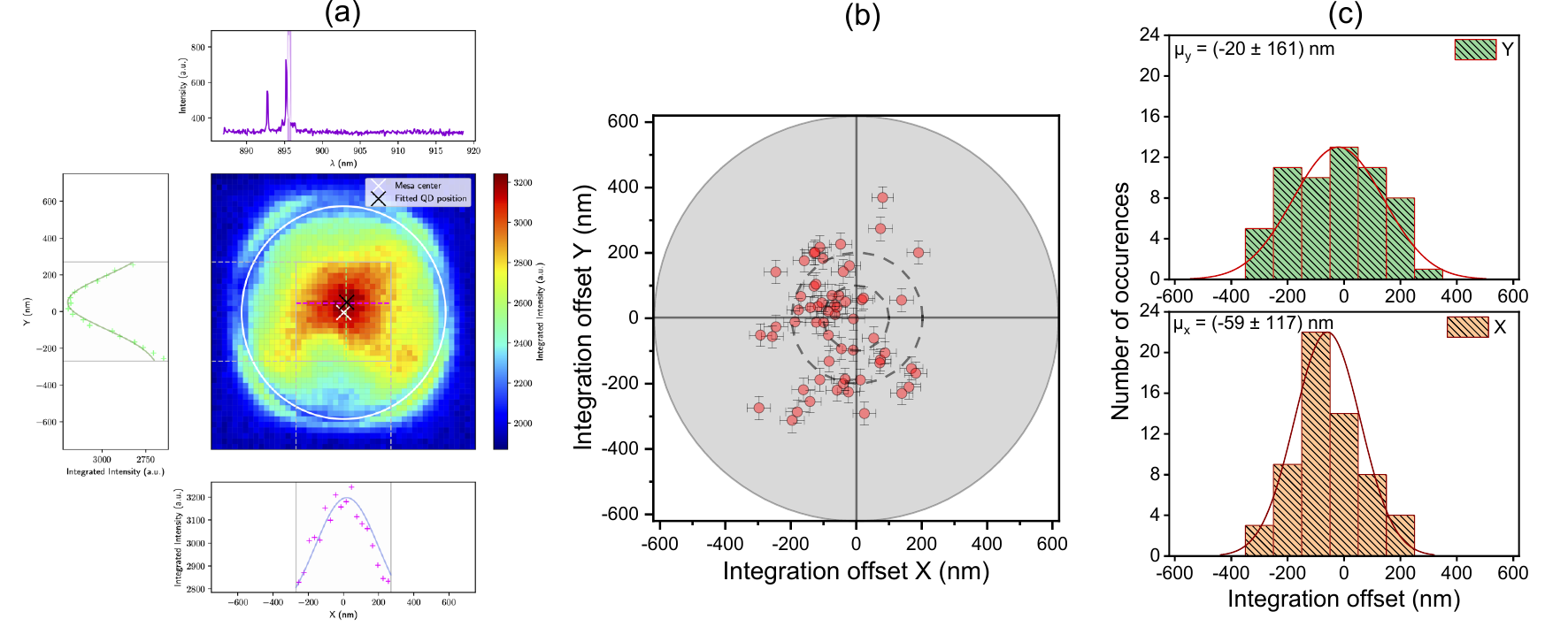}
  \caption {(a) A representative example of a 2D Gaussian function fitted to the integrated emission intensity (in the indicated spectral range of 895.4 to 895.9 nm) of the QD CL spectra. This fitting is used to extract the emission center of the deterministically integrated QD with an offset of $(66 \pm 34)$ nm. (b) Spatial distribution of determined QD positions of a total of 60 deterministically fabricated QD devices in the 2D plane accompanied by error bars derived from fit residuals and SEM scan resolutions, defined as the vectorial distance between the QD emission center (from the Gaussian fit of CL emission) and the geometric center of the mesa structure (from SEM analysis), The gray area indicates the size of the mesa. Dashed circles with a diameter of 100 nm and 200 nm contain approximately 20\% and 55\% of the devices, respectively. (c) The offset histograms along the x and y axes were then fitted with Gaussian functions, yielding a mean offset of $\mu_x = (-59 \pm 117)$ nm and $\mu_y = (-20 \pm 161)$ nm. These values represent the upper-bound alignment precision and accuracy due to extended and asymmetric CL emission profiles.}
  \label{fig:3}
\end{figure}

The precise spatial alignment of quantum emitters relative to their surrounding photonic structures is crucial for achieving high-performance single-photon sources \cite{Rickert2025}. A two-step characterization protocol combining CL and SEM imaging was implemented to quantitatively evaluate the alignment accuracy of the single-step iEBL integration process. After the completion of the device fabrication, CL mapping was performed at 20 K temperature with 3 kV acceleration voltage and a minimal aperture of 20 µm to identify the emission centers of the QDs. Each CL map was obtained from a field of view that was precisely centered on the mesa of the fabricated CBG structures. The acquisition of the emission was recorded with high spectral and spatial resolution in order to quantify the features of the individual QDs.

The QD position was extracted for each structure by fitting a 2D Gaussian function to a subregion of the CL intensity map using a least-squares optimization method. The region of interest was manually selected to encompass the area of maximum integrated intensity, where the QD signal predominates. This subarea is illustrated as a gray rectangle in Fig. \ref{fig:3}(a). 

However, for the current CBG geometries, determining the QD emission center is inherently more prone to error than for larger photonic structures \cite{madigawaDonges2024, Gshrey2015}. The spatial extent of the CL emission spots is larger than the extent of the mesas, resulting in a non-perfect Gaussian profile within the region of interest. The extended profile is likely attributed to the long carrier diffusion lengths observed in the high-quality epitaxial layers and hybrid bonded interface. Furthermore, the CL spots frequently exhibit deviations from circular symmetry, likely due to microscopic variations in the bonding interface that locally affect carrier diffusion and recombination. Such asymmetries, which are visible in the CL maps, have the potential to shift the fitted deviation ($\Delta r_{QD}$) from the true QD position.

Parallel to each CL mapping, a high-resolution SEM image of the same device, covering a larger area than the mesa, was utilized to extract the mesa radius and geometric center with nanometer-scale precision. The extracted edge of the mesa structure is depicted as a white circle on the CL map in Fig. \ref{fig:3}(a). Nonetheless, the process of identifying the mesa center in SEM images is subject to additional deviations due to factors such as angled detector geometry and variations in side wall contrast, which are reflected in the given uncertainties. 

Figure. \ref{fig:3}(b) illustrates the nominal mesa diameter as a reference for the device footprint, represented by the gray circular area. The red dots indicate the QD positions extracted from Gaussian fits of the CL maps. The error bars represent the combined uncertainty from maximum fit residuals and SEM pixel resolution. The inner dotted circle has a diameter of 100 nm and contains about 20\% of the devices. The outer dotted circle has a diameter of 200 nm and encloses about 55\% of the devices.

As can be seen in Fig. \ref{fig:3}(c), the histograms of integration offset along the X and Y axes. Gaussian fits to the data yield mean offsets of $\mu_x = (-59 \pm 117)$ nm and $\mu_y = (-20 \pm 161)$ nm, indicating a slightly larger variation along the X-axis.

These values are comparable to, though slightly larger than, those reported previously for iEBL of larger photonic structures \cite{madigawaDonges2024}, reflecting the higher relative difficulty of position determination for the QDs in smaller mesas with the large diffusion length of the present work. The asymmetry in $\mu_x$ and $\mu_y$ is likely to originate from a directional mechanical drift of the cryostat's cold finger during the iEBL process. Together, these findings quantitatively substantiate the alignment performance and provide an upper limit estimate of the integration accuracy for this device class.
The alignment accuracy can be further improved by optimizing the EBL writing strategy. For instance, in the present workflow, the CBG is patterned in a conventional meander mode, where the exposure begins at the bottom-left corner of the write field and proceeds line by line in the upward direction across the field. As a result, the central mesa region containing the QD is written only after a finite delay of up to 3 minutes, during which random thermal drift of the cold stage, typically in a range of 5-10 nm per minute, can displace the relative QD position, leading to comparatively larger misalignment. A more robust approach for future device generations is to initiate the iEBL writing directly from the mesa center, thereby minimizing the time delay between alignment and pattern exposure and significantly reducing the probability of drift-induced misalignment.

\section{Impact of ring number and alignment accuracy on brightness} 

To evaluate the scalability and optimization potential of our QD-CBG devices further, we investigated how the PEE depends on $\Delta r_{QD}$ and the number of Bragg rings. The PEE values were obtained for 83 out of 95 optically active devices.
These parameters are crucial for balancing optical performance with fabrication complexity. This section compares the simulated extraction trends with actual experimental results and examines how structural complexity influences device yield and brightness across different designs. 

\begin{figure}
 \centering
  \includegraphics[width=0.9\linewidth]{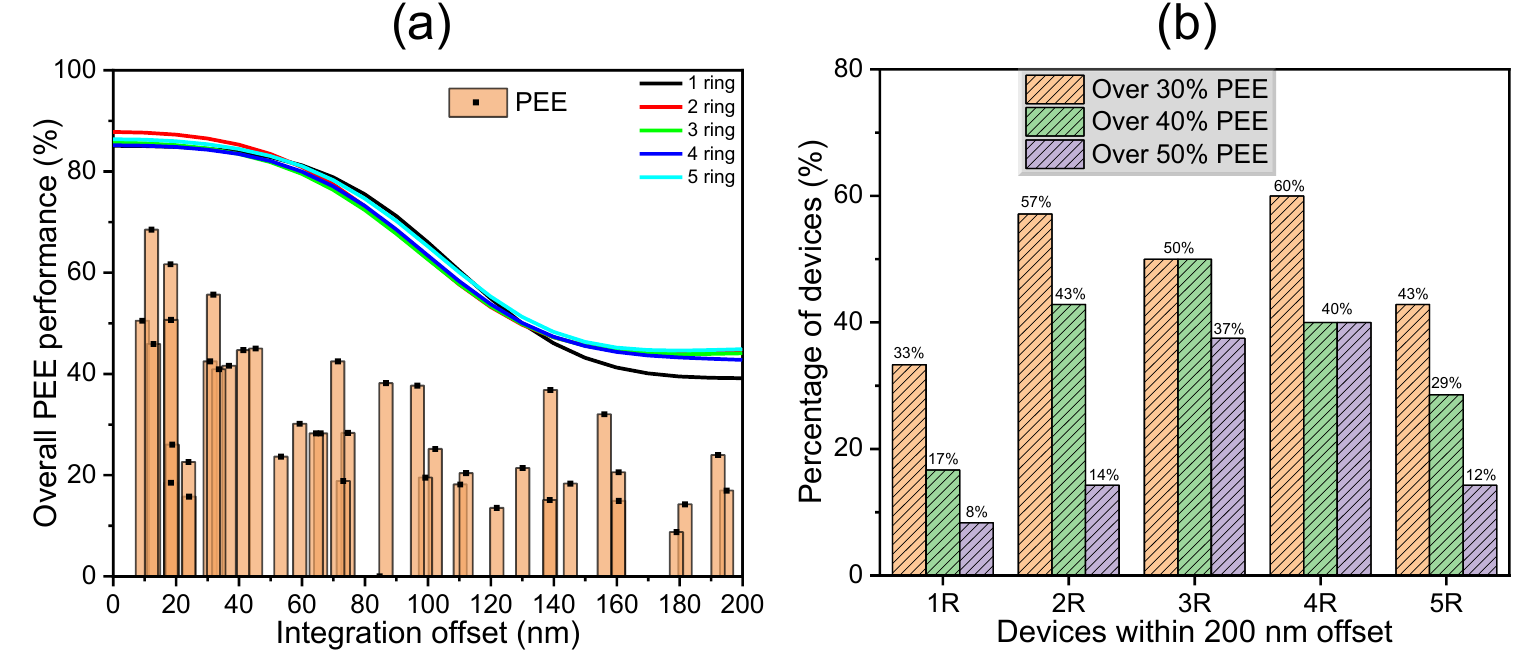}
  \caption{ (a) Photon extraction efficiency versus $\Delta r_{QD}$. The simulated data (lines) for 1-ring (black) to 5-ring (cyan) devices are combined with experimentally measured PEE values (bars). (b) Statistical distribution of devices exhibiting over 30\%, 40\%, and 50\% is grouped by CBG geometry and evaluated within a 200 nm integration offset. The employment of 3-4 ring designs has been demonstrated to consistently yield high-performing devices, thereby highlighting their suitability for scalable quantum photonic applications.
}
  \label{fig:4}
\end{figure}

Figure \ref{fig:4}(a) presents the PEE values of the devices as a function of $\Delta r_{QD}$. The continuous lines represent FEM simulation results for various Bragg ring configurations, while the bar plot shows experimentally measured PEE values obtained from a batch of devices, derived from the estimated setup efficiency of 7.3\% (See supporting information for more details). A strong correlation between experiment and simulation validates the model and confirms that PEE degrades significantly with increasing $\Delta r_{QD}$.

Our statistical evaluation reveals that at least 20\% of the fabricated QD-CBG devices achieve a PEE exceeding 40\% when the spatial integration offset is below 200 nm.
A similar trend is observed in the simulated value of PEE, which remains largely independent of the number of rings. The geometry was first optimized for a single-ring CBG and scaled for higher rings, considering fabrication tolerances, with critical parameters held constant. In this specific CBG design, it was observed that a larger mesa preserves high extraction for lateral QD offsets up to $\pm200$ nm.  Notably, the experimentally measured PEE is consistently lower than the simulated predictions, which can be attributed to practical fabrication imperfections such as structural deviation from the device geometries, imperfect sidewalls, non-ideal etching profiles, and a non-ideal internal quantum efficiency of the QDs \cite{reitzenstein2010quantum,grosse2021quantum}. 
Figure \ref{fig:4}(b) shows the statistical distribution of functional devices, categorized by the number of Bragg rings and grouped by performance threshold ($>30\%$, $>40\%$, and $>50\%$ PEE), evaluated within a 200 nm radial misalignment window. The results indicate that the 3- and 4-ring designs yield the highest average percentage of high-performance devices, achieving a favorable balance between fabrication simplicity, integration tolerance, and optical performance. This behavior is attributed to the aforementioned fabrication-induced limitations, where a larger number of rings is required to achieve optimal performance and to compensate for the associated loss channels. However, as illustrated in Figure \ref{fig:4}(b), continuously increasing the number of rings is not necessarily advantageous. For high ring counts, such as 5 rings, the integration process to write patterns becomes more time-consuming and is more susceptible to spatial misalignment arising from a random stage drift. Consequently, the number of rings should be carefully optimized to balance performance enhancement against fabrication complexity and alignment tolerances. 

These results strongly motivate optimizing the CBG architecture toward compact geometries incorporating only 3 to 4 Bragg rings, wherein the present case, the PEE remains within $\sim80-90\%$ of the maximum achievable value. Such designs not only maintain robust photon extraction but also reduce fabrication complexity, shorten lithography write times by more than a factor of two, and minimize the device footprint for dense integration. The reduced dimension further enhances compatibility with ring-based electrically controlled platforms \cite{Setthanat2024, mudiBarua2025} and enables scalable, high-density on-chip photonic circuit architectures without compromising optical performance.  

\section{Quantum optical performance assessment}

A systematic optical characterization was performed on a subset of 41 QD-CBG devices across multiple fabrication batches to evaluate their optical and quantum optical performance. For this purpose, the sample was mounted in a closed-cycle helium (He) cryostat with a base temperature of 4 K (see SI for setup details). The devices were quasi-resonantly excited using a mode-locked optical parametric oscillator (OPO) laser tuned to 881 nm. We analyzed the emission properties through time-integrated spectra, time-resolved PL (lifetime), second-order autocorrelation \(g^{(2)}(0)\) for single-photon purity, and Hong-Ou-Mandel (HOM) two-photon interference to assess photon indistinguishability.

\begin{figure}
 \centering
  \includegraphics[width=1\linewidth]{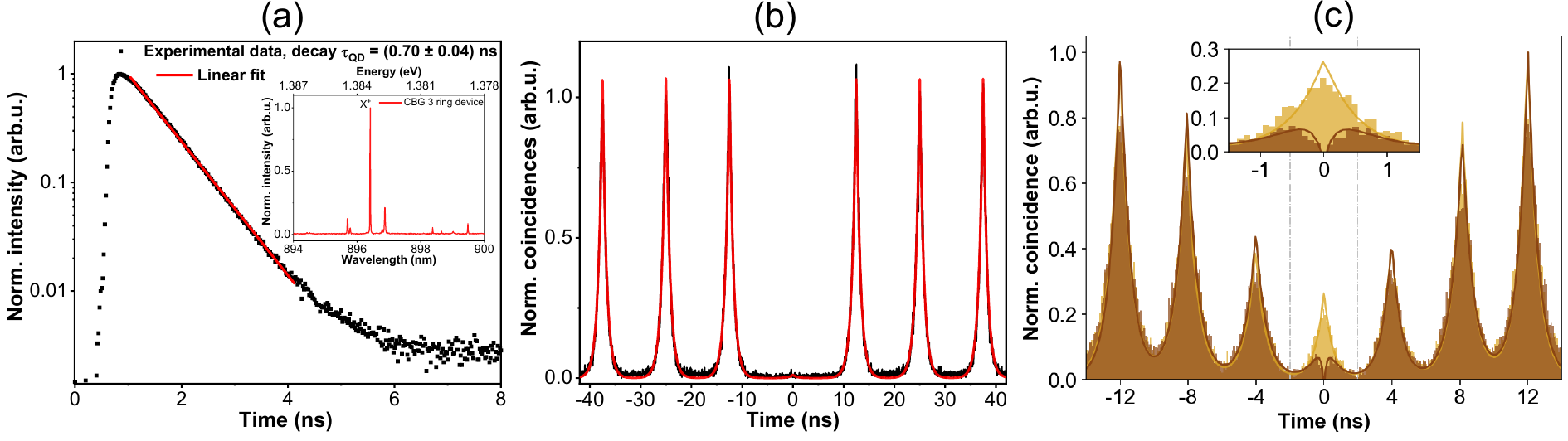}
  \caption{Time-resolved and quantum optical characterization of a representative 3-ring QD–CBG device. (a) Time-resolved PL decay under 881 nm pulsed excitation, fitted linearly, yielding a spontaneous emission lifetime of \(\tau_{\mathrm{QD}}=(0.70 \pm 0.04)\) ns (inset: spectrum shows the charged exciton ($X^{+}$) emission). (b) Second-order autocorrelation histogram showing \(g^{(2)}(0)=0.011 \pm 0.002\), indicating high single-photon purity, and (c) HOM interference histogram with a 4 ns pulse delay, revealing raw visibility of 53\% and fitted visibility of 63\% (inset: shows a close-up of the central dip near zero delay).
}
  \label{fig:5}
\end{figure}

Figure~\ref{fig:5} presents the time-resolved and quantum optical characterization of a representative device featuring a 3-ring CBG. Fig,~\ref{fig:5}(a) shows the time-resolved (PL) trace from a QD-CBG under pulsed excitation. The corresponding spectrum, shown in the inset, reveals resolution-limited emission lines, with the charged exciton transition emerging as the dominant feature. The PL signal exhibits a fast initial decay with a fitted lifetime of $(0.70 \pm 0.04)$ ns. For comparison, planar QDs without cavity coupling typically exhibit lifetimes of $\sim$ $(1.09 \pm 0.04)$ ns, indicating an almost negligible Purcell enhancement in the present device. The emission trace of QD-CBG also exhibits a weak, long-lived tail. This extended component has previously been attributed to ultra-slow carrier capture mechanisms, such as Auger- or phonon-assisted transitions, which can result in charge carrier trapping and delayed recombination events \cite{Nguyen2013, Arnold2014}. Such processes may induce fluctuations in the QDs' charge state, thereby affecting both the temporal coherence and timing stability of the emitted photons. Fig, ~\ref{fig:5}(b) displays the second-order correlation histogram measured under pulsed excitation. A measured value of $g^{(2)}(0) = 0.011 \pm 0.002$ indicates a strong suppression of multi-photon events, confirming highly pure single-photon emission from the device.

\begin{table}[h!]
\centering
\caption{Lifetime and $g^{(2)}(0)$ for different device geometries. Representative device performance from each configuration is presented with (uncertainty).}
\resizebox{0.85\textwidth}{!}{
\begin{tabular}{|c|c|c|c|c|c|c|}
\hline
\textbf{Device geometry} & \textbf{1-R} & \textbf{2-R} & \textbf{3-R} & \textbf{4-R} & \textbf{5-R} & \textbf{Planar QD} \\
\hline
\textbf{Lifetime (ns)} 
& $0.95 \pm 0.03$ & $1.05 \pm 0.03$ & $0.70 \pm 0.04$ & $0.71 \pm 0.04$ & $0.71 \pm 0.03$ & $1.09 \pm 0.04$ \\
\hline
\(\boldsymbol{g^{(2)}(0)}\) 
& $0.072 \pm 0.002$ & $0.048 \pm 0.003$ & $0.011 \pm 0.002$ & $0.072 \pm 0.004$ & $0.029 \pm 0.003$ & $0.18 \pm 0.04$ \\
\hline
\end{tabular}
}
\end{table}

Figure ~\ref{fig:5}(c) provides the HOM interference histogram obtained under the same excitation condition. A temporal delay of 4 ns was introduced between successive excitation pulses to enable two-photon interference between sequentially emitted photons. The experiment yielded a raw visibility of approximately \(53(2)\%\) and a fitted visibility of \(63(1)\%\). These values are primarily limited by dephasing mechanisms and time jitter inherent to an imperfect excitation scheme \cite{Somaschi2016, Ding2016}. Notably, the HOM interference visibility could be significantly improved by employing resonant techniques such as phonon-assisted excitation or strictly resonant \(\pi\)-pulse excitation \cite{REITZENSTEIN2025689}.

The combined analysis of measured high single-photon purity, sub-nanosecond radiative lifetime, and moderate photon indistinguishability demonstrates that our single-step, \textit{in-situ} EBL process reliably produces high-quality quantum light sources with excellent reproducibility. Table 2 shows the representative device performances on charge-exciton lifetime and second-order correlation values for devices with different ring geometries. While the $g^{(2)}(0)$ remains largely independent of the ring number, the lifetime decreases with increasing ring numbers, reflecting a slight Purcell enhancement, as it was found in simulations. Additional single-photon performance metrics across devices and field regions are provided in the SI. These results validate the robustness of our fabrication approach across different device geometries and multiple production batches. They reinforce the scalability of the integration method by showing that deterministic emitter–cavity coupling can be achieved without degradation of optical performance.

\section{Discussion}

\begin{table}[h!]
\centering
\caption{Optical performance comparison with state-of-the-art devices. \linebreak
Abbreviations: $\Delta R$: Alignment accuracy, PEE: Photon extraction efficiency, V: Photon indistinguishability, N: Number of functional devices, I.P: Integration procedure, WL: Wavelength, Ref: Reference, MB: marker-based (multiple step), iEBL: In-situ electron beam lithography (single step), R: Resonant, QR: Quasi-resonant, NR: Non-resonant, TPE: Two-photon excitation, DBR: Distributed Bragg reflector.}
\resizebox{\textwidth}{!}{
\begin{tabular}{|c|c|c|c|c|c|c|c|c|}
\hline
\textbf{Cavity structure} & \textbf{$\Delta R$ (nm)} & \textbf{PEE (\%)} & \textbf{V} & \textbf{N} & \textbf{I.P } &\textbf{WL (nm)} & \textbf{Excitation} & \textbf{Ref.} \\
\hline

Mesa and hybrid-CBG  & 30-100  & --  & --  &--  & MB/iEBL&900-910 &--&\cite{madigawaDonges2024}   \\
Mesa                 & 50      & --  & --  & 3  &iEBL    &930-937 &NR& \cite{gschrey2013situ}         \\
Pillar with rings    & 50-100  & 39  & 0.21& 74 &MB      &775-785 &QR&\cite{madigawa2025}\\
Hybrid-CBG& --      & --& 0.96 & 4 &MB &900-940&R& \cite{Rickert2025}\\
Hybrid-CBG& 35      & --& -- & 10 &MB &784-806&TPE& \cite{krieger2024postfabrication}\\
Hybrid-CBG& 100      & --& -- & 14 &MB &793-796,918-926&--& \cite{peniakov2024polarized}\\
Hybrid-CBG           & $10-200$      & 68  & 0.63 & $95$ &iEBL    &890-900 &QR&This Work \\
\hline
\end{tabular}
}
\end{table}

Table 3 benchmarks the performance of state-of-the-art circular mesa-based devices fabricated via EBL against prior reports, highlighting the critical role of spatial alignment accuracy in determining device efficiency and reproducibility. Collectively, these results establish cryogenic single-step iEBL as a reliable and scalable technique for site-selective integration of QDs into CBG resonators, offering high yield, spectral stability, and structural uniformity across dense device arrays and diverse geometries, while also shedding light on its practical limitations and optimization potential. By avoiding the need for alignment markers and thermal cycling, 95 out of the 103 fabricated devices were found to be optically active, reflecting the robustness of the process. A hybrid analysis combining CL-based Gaussian fitting and SEM imaging further reveals that at least 41 devices have alignment accuracies within $\Delta r_\text{QD}< 200$ nm. This comparatively modest alignment accuracy, however, reflects inherent challenges such as carrier diffusion and random thermal drift, which limit the exact determination of QD positions in the preselection and lithography processes. Despite these constraints, the method continuously yields devices with high brightness and experimental PEE values reaching up to 68\%. Quantum optical characterization of the devices confirms reproducible performance on key single-photon metrics. A decay time of $(0.70 \pm 0.04)$ ns was observed in the lifetime measurements of the charged exciton. The utilization of autocorrelation measurement facilitates the confirmation of single-photon purity. The average value of $g^{(2)}(0)$ from 1-5 ring devices was calculated to be \(0.046 \pm 0.027\). Furthermore, HOM interference under quasi-resonant excitation resulted in raw and fitted visibilities of \(53(2)\%\) and \(63(1)\%\), respectively, from the representative device. As shown in Table 3, these values correspond to or exceed those previously reported, while also offering the advantage of high fabrication yield for scalable applications.
An equally important aspect of this study is the geometry-dependent analysis of the CBG design. Both the experiment and FEM simulations reveal the PEE value saturates around 3-4 ring devices, with an average theoretical value reaching up to $(86.0 \pm 1.4)$\%. Although a higher PEE value (96\%) is reported \cite{yao2018design} by employing smaller mesas and more number of rings at 930 nm emission, such a configuration lies beyond the scope of this study, as the geometries of the devices used in this study (1-5 rings) were maintained to ensure consistent comparison. Beyond 4 rings, the increase in fabrication footprint and susceptibility to thermal drift outweighs any marginal performance gain.
Importantly, further improvement in alignment accuracy is anticipated by modifying the EBL writing sequence to initiate patterning from the mesa center, thereby reducing drift between alignment and exposure. These findings demonstrate that high-yield, compact, and robust device architectures can be reliably achieved, providing a practical pathway toward large-scale integration of deterministic single-photon sources.

\section{Conclusion}

The iEBL nanotechnology platform has been used and statistically evaluated for the deterministic integration of InGaAs QD into CBG resonators with 10-200 nm accuracy, acting as bright single-photon sources. The experimental PEE value for the ideal device reaches as high as 68\%. Further, this method attains over 90\% process yield while eliminating the necessity for alignment markers.
Comprehensive structural and optical characterization confirms high uniformity and reproducible single-photon performance across large device batches, as observed for all 95 optically active devices. It is of particular significance that CBG designs of 3-4 Bragg rings offer optimal photon extraction with minimal complexity and footprint, rendering them well-suited for faster, denser integration with the prospects of ridge-based electrically controlled devices.
These results establish iEBL as a robust platform for high-throughput fabrication of scalable quantum light sources, enabling future on-chip quantum optical technologies.

\medskip

\medskip
\textbf{Acknowledgements} \par 
The authors acknowledge financial support from: DFG Project RE2974/28-1 (448532670); German Ministry of Science and Education (BMBF) Project QR.N (16KIS2193).
The authors at KIST (SIP and JDS) acknowledge the partial support from the institutional program of KIST and global top program.

\medskip
\textbf{Data availability} \par 
All data supporting the findings of this study are available within the manuscript and its supplementary information. Additional data are available from the corresponding author upon reasonable request.

\medskip
\textbf{Competing interest} \par
The authors have declared that there are no competing interests.

\medskip

\bibliographystyle{unsrt}
\bibliography{Reference}

\end{document}